\documentstyle[twoside]{article}

\catcode`\@=11
\long\def\@makefntext#1{
\protect\noindent \hbox to 3.2pt {\hskip-.9pt  
$^{{\eightrm\@thefnmark}}$\hfil}#1\hfill}

\def\@makefnmark{\hbox to 0pt{$^{\@thefnmark}$\hss}}	
	
\def\ps@myheadings{\let\@mkboth\@gobbletwo
\def\@oddhead{\hbox{}
\rightmark\hfil\eightrm\thepage}   
\def\@oddfoot{}\def\@evenhead{\eightrm\thepage\hfil
\leftmark\hbox{}}\def\@evenfoot{}
\def\sectionmark##1{}\def\subsectionmark##1{}}

\oddsidemargin=\evensidemargin
\newcounter{sectionc}\newcounter{subsectionc}\newcounter{subsubsectionc}
\renewcommand{\section}[1] {\vspace{12pt}\addtocounter{sectionc}{1} 
\setcounter{subsectionc}{0}\setcounter{subsubsectionc}{0}\noindent 
	{\tenbf\thesectionc. #1}\par\vspace{5pt}}
\renewcommand{\subsection}[1] {\vspace{12pt}\addtocounter{subsectionc}{1} 
	\setcounter{subsubsectionc}{0}\noindent 
	{\bf\thesectionc.\thesubsectionc. {\kern1pt \bfit #1}}\par\vspace{5pt}}
\renewcommand{\subsubsection}[1] {\vspace{12pt}\addtocounter{subsubsectionc}{1}
	\noindent{\tenrm\thesectionc.\thesubsectionc.\thesubsubsectionc.
	{\kern1pt \tenit #1}}\par\vspace{5pt}}
\newcommand{\nonumsection}[1] {\vspace{12pt}\noindent{\tenbf #1}
	\par\vspace{5pt}}

\newcommand{\textlineskip}{\baselineskip=13pt}

\def\eightcirc{
\begin{picture}(0,0)
\put(4.4,1.8){\circle{6.5}}
\end{picture}}
\def\eightcopyright{\eightcirc\kern2.7pt\hbox{\eightrm c}}

\def\abstracts#1#2#3{{
	\centering{\begin{minipage}{4.5in}\baselineskip=10pt\footnotesize
	\parindent=0pt #1\par 
	\parindent=15pt #2\par
	\parindent=15pt #3
	\end{minipage}}\par}} 

\newcommand{\bibit}{\nineit}

\renewenvironment{thebibliography}[1]
	{\frenchspacing
	 \ninerm\baselineskip=11pt
	 \begin{list}{\arabic{enumi}.}
	{\usecounter{enumi}\setlength{\parsep}{0pt}
	 \setlength{\leftmargin 12.7pt}{\rightmargin 0pt} 
	 \setlength{\itemsep}{0pt} \settowidth
	{\labelwidth}{#1.}\sloppy}}{\end{list}}

\newcounter{itemlistc}
\newcounter{romanlistc}
\newcounter{alphlistc}
\newcounter{arabiclistc}

\def\@citex[#1]#2{\if@filesw\immediate\write\@auxout
	{\string\citation{#2}}\fi
\def\@citea{}\@cite{\@for\@citeb:=#2\do
	{\@citea\def\@citea{,}\@ifundefined
	{b@\@citeb}{{\bf ?}\@warning
	{Citation `\@citeb' on page \thepage \space undefined}}
	{\csname b@\@citeb\endcsname}}}{#1}}

\newif\if@cghi
\def\cite{\@cghitrue\@ifnextchar [{\@tempswatrue
	\@citex}{\@tempswafalse\@citex[]}}
\def\citelow{\@cghifalse\@ifnextchar [{\@tempswatrue
	\@citex}{\@tempswafalse\@citex[]}}
\def\@cite#1#2{{$\null^{#1}$\if@tempswa\typeout
	{IJCGA warning: optional citation argument 
	ignored: `#2'} \fi}}

\def\pmb#1{\setbox0=\hbox{#1}
	\kern-.025em\copy0\kern-\wd0
	\kern.05em\copy0\kern-\wd0
	\kern-.025em\raise.0433em\box0}

\def\fnt#1#2{\footnotetext{\kern-.3em
	{$^{\mbox{\scriptsize #1}}$}{#2}}}

\def\fpage#1{\begingroup
\voffset=.3in
\thispagestyle{empty}\begin{table}[b]\centerline{\footnotesize #1}
	\end{table}\endgroup}

\def\runninghead#1#2{\pagestyle{myheadings}
\markboth{{\protect\footnotesize\it{\quad #1}}\hfill}
{\hfill{\protect\footnotesize\it{#2\quad}}}}
\font\tenrm=cmr10
\font\tenit=cmti10 
\font\tenbf=cmbx10
\font\bfit=cmbxti10 at 10pt
\font\ninerm=cmr9
\font\nineit=cmti9

\font\eightrm=cmr8

\newfont{\msbm}{msbm10}

\def\qed{\hbox{${\vcenter{\vbox{			
   \hrule height 0.4pt\hbox{\vrule width 0.4pt height 6pt
   \kern5pt\vrule width 0.4pt}\hrule height 0.4pt}}}$}}

\def\Z{\hbox{\msbm Z}}

\def\R{\hbox{\msbm R}}

\def\be{\begin{equation}}

\def\ee{\end{equation}}

\def\f{\phi}

\def\wh{\widehat}

\def\ri{{\rm i}}

\def\h2{${\rm h}(2)$}
\begin{document}

\runninghead{Comments on a Full Quantization of the Torus}
{Comments on a Full Quantization of the Torus}

\normalsize\textlineskip
\thispagestyle{empty}
\setcounter{page}{1}


\vspace*{0.88truein}

\fpage{1}
\centerline{\bf COMMENTS ON A FULL QUANTIZATION OF THE TORUS}
\vspace*{0.37truein}
\centerline{\footnotesize J.M. VELHINHO}
\vspace*{0.015truein}
\centerline{\footnotesize\it Universidade do Algarve, U.C.E.H., 
\'Area Departamental 
de F\'\i sica.}
\baselineskip=10pt
\centerline{\footnotesize\it Campus de Gambelas,
8000 Faro, Portugal}


\vspace*{0.225truein}


\vspace*{0.21truein}

\abstracts{
Gotay showed that a representation 
of the whole Poisson algebra
of the torus given by geometric quantization
is irreducible with respect to the most natural 
overcomplete set of observables. We study this representation
and argue that it cannot be considered as 
physically acceptable. In particular, classically
bounded observables are quantized by operators with unbounded
spectrum. Effectively, the latter amounts to
lifting the constraints that compactify both directions in the torus.}{}{}

\vspace*{1pt}\textlineskip	
  
\vspace*{-0.5pt}




\section{Introduction}
\noindent
There are still some open questions as to
what is precisely meant by quantization of 
a general symplectic manifold. 
Two interconnected problems affect Dirac's original formulation:
(a) which Poisson subalgebra should one choose to impose 
the Dirac quantum condition ``Poisson bracket goes
to commutator''; (b) which extra conditions should be required in
order to guarantee that the quantum representation 
is physically meaningful.

Although no general result was ever proved, it is a common belief that 
a physically acceptable quantization
of the whole Poisson algebra cannot be found, 
for any symplectic manifold (see below).
One way to circumvent the obstruction to a full quantization
consists in looking for quantum
representations of some convenient proper subalgebra $\cal S$ of the whole
Poisson algebra.
It is generally acknowledged that the subalgebra $\cal S$ to be quantized 
must be (kinematically) complete, in the sense that
it should contain a complete set of
classical observables. However, the precise definition of a complete set of
classical observables  $\cal B$, $\cal B\subset\cal S$, is not so consensual.
In Ref.~1, A.A.~Kirillov defines a set of observables $\cal B$ to be complete
if {\em the only observables which (Poisson) commute with every element of}
 $\cal B$ {\em are the constant functions} (this, in particular, means that
the functions in $\cal B$  separate points locally almost everywhere on 
the phase space). A stronger condition (in the sense that it guarantees local
separation of points everywhere) is required by M.J.~Gotay, H.B.~Grundling
and G.M.~Tuynman, reading: {\em the Hamiltonian vector fields associated with
the functions in $\cal B$ should generate the tangent space everywhere in 
phase space}.\cite{G-G-T}
Notice that, as the authors acknowledge, none of the above two conditions
guarantees global separation of points.
The latter, we think, must be imposed in order to make 
sure that one 
is quantizing the initial phase space and not 
its quotient by the relations
defined by considering two points equivalent 
if they are not separated by the 
functions in the set $\cal B$. A similar strong notion of completeness
is favored by A.~Ashtekar, which requires that {\em the subalgebra}
$\cal S$ {\em should be ``large enough'' so that any (sufficiently regular)
function on the phase space can be represented as (possibly a limit of)
a sum of products of elements of $\cal S$}.\cite{As2}

In order to avoid the appearance at the quantum level
of spurious degrees of freedom,
one imposes, in one way or another, irreducibility conditions on 
the quantum representations
(see Ref.~2 for a discussion). Following Ref.~2 (see also Ref.~1), 
we will concentrate on an explicit  
irreducibility requirement: 
{\em the quantum Hilbert space should be irreducible with respect to the
action of all operators representing a complete set of classical observables}.
 
Many examples of representations of ``large'' 
Poisson subalgebras (possibly the whole algebra) which do not comply with
irreducibility are known; for instance the so called {\em prequantizations}
obtained through the methods of Geometric Quantization.\cite{Ki}
It is generally acknowledged that
these examples are manifestations of a general obstruction to quantization:
the widely accepted ``no-go conjecture'' claims, in essence, that a 
physically relevant quantization of a full Poisson algebra is unattainable, 
due to the impossibility of satisfying simultaneously the Dirac quantum
condition on the whole algebra and the irreducibility requirement.
The first rigorous ``no-go'' result goes
back to the works of Groenewold and Van Hove (see 
Ref.~2 and references therein) and states 
that it is impossible to fulfill the Dirac quantum condition
for the whole Poisson algebra on $\R^2$
while keeping irreducibility on the level of the Poisson 
subalgebra generated by $q$ and $p$ 
(the Heisenberg algebra ${\rm h}(2)={\rm span}\{1,q,p\}$ of inhomogeneous 
degree one polynomials
in $q$ and $p$). 
This is the famous ``no-go theorem''
for $\R^{2}$ (which is readily generalizable to $\R^{2n}$). 
Recently, similar no-go
results also based on irreducibility were proven for the
sphere $S^2$ and the cylinder $T^{\ast}S^1$.\cite{G-G-H,G-G}

The standard quantization procedure requires therefore the selection of
a preferred subalgebra $\cal S$ of classical observables.
This special subalgebra must be complete and  
should be such that irreducibility 
can be achieved. 
It is also assumed that 
{\em the constant function $1$ belongs to $\cal S$ and should be 
mapped to the identity operator}
(see Ref.~1 and Ref.~2). Clearly, no generality is lost in assuming
that $1\in \cal S$, for given a subalgebra $\cal S$ which does not 
contain $1$, every acceptable
quantization map can be extended just by assigning the identity 
operator to the function $1$. 
Notice also that the above
condition on the quantization of $1$ is independent from irreducibility 
and is of a quite different nature.
The ``$1$ goes to $\bf 1$'' condition is 
required to enforce the correspondence between the classical value 
attained by the constant observable and its quantum spectrum.
 
Unlike the above outlined conditions, when considering general observables 
there seems to be no major
agreement on the important subject of the relation between the classical 
range of values attained and the 
quantum spectrum of the observable.
This is closely related with the
issue of preservation of the classical multiplicative structure. 
Along with the Poisson structure, pointwise multiplication of
observables is a fundamental aspect of the classical theory and 
one certainly would like to see it reflected at the quantum level.
The question is to what extent should the multiplicative
structure be preserved upon quantization, given that
the Poisson and the multiplicative structure
are not fully compatible at the quantum level (see Ref.~2).
Due to this fact, some authors take the point
of view that quantization concerns the preservation of Poisson relations 
only, thus choosing not to impose {\it a priori} any further conditions
on quantization (besides irreducibility and 
``$1$ goes to $\bf 1$'').\cite{G-G-T}
On the other hand, one can find in the literature explicit concern about
the relations between the classical range (which we will call classical 
spectrum) and the quantum spectrum (Ref.~6) and also the
requirement that relations among preferred classical observables 
coming from the multiplicative structure should
be implemented in the quantum theory through appropriate anti-commutation
relations (Ref.~3, Ref.~7).

In our opinion, multiplicative algebraic relations among preferred 
classical observables are fundamental and should be carried over to the 
quantum theory. Let us assume that a preferred 
Poisson subalgebra $\cal S$ of real valued classical observables has been
chosen. Although $\cal S$ is, in general, not closed under multiplication,
it still reflects the 
multiplicative structure and also the global topology of phase space. 
Let us consider first $\R^2$. 
In this case, global, canonically conjugate, coordinates $q$ and $p$ 
exist and the algebra \h2 generated by these observables is complete. 
Of course, no
relations can be found among the generators of \h2,
since $q$ and $p$ are independent, linearly or otherwise. A different 
situation occurs if the Poisson subalgebra 
$\,{\rm span}\{1,q,q^2,p,p^2\}$ of polynomials
of degree no greater than two in $q$ and $p$ is chosen. In this case, 
multiplication relates, for instance, $q$ and $q^2$, encoding the obvious fact
that $q^2$ is a positive valued observable whose values in every state are
completely determined by the observed values of $q$. 

Whereas in $\R^2$ (or $\R^{2n}$) it is always possible to choose a complete 
Poisson subalgebra which is free of relations,
this is no longer the case
when the phase space has a globally non-trivial topology. In the absence
of a global chart, a set of everywhere defined functions which separates
points is necessarily overcomplete, meaning that the number of functions
in the set exceeds the dimensionality of the phase space. Clearly, the 
functions in a  overcomplete set cannot be all independent and therefore
relations among them must appear. A standard example of this situation
is provided  by the cylinder $T^{\ast}S^1$, the phase space of
a particle moving on a circle. Since the
angular variable $\theta$ on $S^1$ is not globally
defined, the best one can do is to work with the functions $\sin\theta$,
$\cos\theta$ and $p$, where $p$ is the conjugate momentum of $\theta$.
In general, the global non-triviality of the phase space will therefore 
manifest 
itself in every admissible preferred subalgebra $\cal S$, in the form
of a set of functional relations among elements of $\cal S$. These
relations carry crucial information about the interdependence of 
observables, their spectrum and the global aspects of the phase space.

In order to illustrate the problems that might occur if algebraic relations 
are neglected in the quantization process, let us consider first the
above example of the algebra $\,{\rm span}\{1,q,q^2,p,p^2\}$ in $\R^2$, 
which includes both	
$q$ and $q^2$. Upon quantization, one then has
two well defined commuting observables ${\hat q}$ and ${\widehat {q^2}}$.
Let us assume that ${\widehat {q^2}}$ is not equal to $({\hat q})^2$.  
Then it is  possible that the operator
${\widehat {q^2}}$, the quantum analog of $q^2$, is not positive,
leading to states with negative expectation values. 
In contrast, even though  $({\hat q})^2$ is a well defined operator which
relates to $\hat q$ as expected, it is not clear 
what meaning should one assign to this operator, since the
quantum analog of $q^2$ was already determined.

One can illustrate further the importance of algebraic relations by 
means of an example of a globally non-trivial phase space.\cite{As2,As-T} 
Consider again the phase space $T^{\ast}S^1$. A natural choice for 
$\cal S$ in this case would be the vector space spanned by the functions 
$\{ 1,\sin\theta,\cos\theta,p\}$. Suppose one is given an irreducible 
Lie-representation $\rho$
of $\cal S$ such that the relation 
\be
\label{1.1}
\big(\rho(\sin\theta)\big)^2+\big(\rho(\cos\theta)\big)^2={\bf 1}
\ee
does not hold. It is then
clear that $\theta$ looses its meaning as an angular variable, and that 
one runs into serious problems with the physical interpretation of the
simultaneously measured values of $\rho(\sin\theta)$ and $\rho(\cos\theta)$.
The natural thing to do is to demand 
(\ref{1.1}) to be satisfied (this condition is, of course, indeed 
satisfied in the usual quantization on $L^2(S^1,d\theta)$,
where $\,\hat p=-{\rm i}\hbar{d\over d\theta}\,$ and 
$\wh {\sin\theta},\ \wh {\cos\theta}$
act by multiplication).
This is the point of view taken in Ref.~3 and Ref.~7, where the authors
propose the exact implementation in the quantum theory of 
algebraic relations among (Poisson) commuting elements of the special classical
subalgebra $\cal S$. Also proposed is the
following natural rule to handle relations among non-commuting
elements of $\cal S$: if $F_1,F_2,\ldots F_m$ and $G$ are 
elements of $\cal S$ such that  $F_1 F_2 \dots F_m$=$G$ then one
requires ${\hat F}_{(1}\dots {\hat F}_{m)}=\hat G$ to be satisfied
for the corresponding quantum operators, where the bracket denotes
symmetrization (this rule generalizes in the natural way to linear
combinations of products).\cite{As2,As-T}

The above considerations indicate that one runs into
serious ambiguities if relations among classical
observables included in the preferred subalgebra $\cal S$ are broken 
at the quantum level.
In particular, if relations among
Poisson commuting (unambiguously quantized) classical observables  
are ignored, one allows the appearance 
of quantum operators whose spectrum bears no relation with the 
classical range, thus making the physical interpretation of the
theory quite problematic. 
A well defined procedure to handle algebraic relations is therefore
an important ingredient of quantization.
An important physical example is provided by non-Abelian
theories of connections with a compact gauge group, where the
gauge invariant configuration space is a globally non-trivial 
infinite-dimensional manifold.\cite{A-I,A-L}
In this case, a convenient (over)complete set of variables is provided
by the Wilson loop variables, but one must deal with the 
algebraic relations among them
(the so-called Mandelstam identities);
this is just the price to pay for working with gauge invariant 
functions (see Ref.~3 and references therein).

A new question which arises in this framework consists in verifying 
the compatibility between the above rule to handle
algebraic relations, on one hand, and the preservation of Poisson 
structure and 
irreducibility, on the other. 
Although this question seems to be a difficult one, in general, one might
gain some insight through the analysis of simple cases. Consider again
$\R^2$, and choose $\cal S$ to be the Poisson subalgebra of polynomials
of degree no greater than two in $q$ and $p$. It is a trivial matter
to verify that the Schr\"odinger quantization of \h2 
extends to this new algebra by the following new assignments:
$$
\wh {q^2}=q^2\ ,\ \ \wh {qp}=-\ri \hbar(q{d\over dq}+1/2)\ , 
\ \ \wh {p^2}=-\hbar^2{d^2\over dq^2}\, .
$$
Since this is an extension of the representation of \h2, irreducibility
is guaranteed. On the other hand,
it is clear that $\,\wh {q^2}=({\hat q})^2$, $\wh {p^2}=({\hat p})^2\,$
and $\, 2\,\wh {qp}=\hat q\hat p+\hat p\hat q$. Thus, in this particular case, 
one finds no incompatibility. The same happens with the
extension of the Schr\"odinger representation of \h2 to the subalgebra
of $C^{\infty}({\R}^2)$ which is made of inhomogeneous degree one
polynomials in $p$, with arbitrary functions of $q$ as coefficients.
Also, for $T^{\ast}S^1$, the above mentioned
quantization of $\,{\rm span}\{1,\sin\theta,\cos\theta,p\}\,$ 
is extendible to functions linear in $p$ in a similar
fashion, again fulfilling both the irreducibility and preservation of
functional relations requirements (see Ref.~2 for a detailed discussion
of these quantizations, as well as obstructions to further extension).
A less trivial and very interesting case is provided by the sphere $S^2$.
Since the area has the same dimensions as the Planck constant $h$, one can
write the canonical area form $\omega$ on $S^2$ as
$$\omega=(Kh/{4\pi}) \sin\theta d\theta d\phi\, ,$$
where $\theta$, $\phi$ are the usual angular variables and $K$ is a
dimensionless constant which determines the total area. The linearly 
independent functions
$$f_1=K/2\ \sin\theta \cos\phi,\ f_2=K/2\ \sin\theta \sin\phi,
\ f_3=K/2\ \cos\theta$$
obviously separate points and are closed under the Poisson bracket:
\be
\label{1.2}
\{f_i,f_j\}={1\over\hbar}\, \epsilon_{ijk}f_k\, .
\ee
The linear space spanned by $f_1$, $f_2$, $f_3$ and the constant function $1$
is thus a natural 
choice for the special subalgebra $\cal S$. One has, of course, the
following relation among elements of $\cal S$:
\be
\label{1.3}
\sum_i f_i^2-\left(K/2\right)^2=0\, .
\ee
A quantization of $\cal S$ is given by a representation
of $\, su(2)$. The irreducible representations of $\cal S$ are thus
well known, indexed by integers or half-integers $j$, and satisfy
\be
\label{1.4}
\sum_i ({\hat f}_i)^2-j(j+1){\bf 1}=0\, .
\ee
Equation (\ref{1.4}) strongly resembles (\ref{1.3}), thus suggesting that
(\ref{1.4}) should be properly interpreted as the quantum counterpart of 
(\ref{1.3}). Suppose first one ignores (\ref{1.3}), thus requiring
irreducibility only as an extra condition on quantization. One then gets
a countable family of equally good candidates to the quantum theory
associated with the sphere of area $Kh$. Uniqueness is obviously 
not achieved. It is only when one demands further the preservation of
relation (\ref{1.3}) that a unique representation is selected. 
Clearly, the relation  (\ref{1.3}) can only be exactly preserved 
for those  values of $K$ which equal $\sqrt {2j(2j+2)}$ for some integer
or half-integer $j$.
Thus, the point of view of exact implementation of functional relations
leads to the conclusion that a proper quantum theory for the sphere can
only be defined for a discrete set of values of the area. This
``quantization'' of the area should not come as a surprise: it is known
that $S^2$ is the reduced phase space associated with a system of two
identical uncoupled harmonic oscillators constrained to have a definite
value of the total energy. Since, from standard Quantum Mechanics, the
allowed values of the energy of such a system form a discrete set,
one immediately obtains the quantization of the area, because
the area of the sphere is determined by the value of energy 
(see Ref.~7 for a quantization of the sphere along these lines).

In the above examples it was possible to preserve the functional relations
within the special subalgebra $\cal S$ in a way which was not 
incompatible with the irreducibility
requirement. In the case of the sphere, although the irreducible
representations automatically fulfill a relation similar to the classical
one, the explicit requirement of exact matching is necessary to obtain a unique
quantization. The additional condition selects from among the set of all
irreducible representations the one with the dimensionality corresponding
to the given area of the classical phase space. 
In particular, one can see that the spectrum of the quantum operators 
${\hat f}_i$ becomes a subset of the range of the corresponding classical 
observables. Notice also that for large 
quantum numbers $j$ the relation between area of phase space and
dimensionality of Hilbert space is just what one might expect from
semi-classical arguments, since both dimensionality and the 
ratio of the area to the Planck constant $h$
approach the integer $2j$.

\section{The Torus}
\noindent

Recently M.J. Gotay et al.\ reexamined the concept of quantization and looked
for obstructions of the Groenewold-Van Hove type for manifolds other
than  $\R^{2n}$. Their work is presented in a series of papers which,
besides formulating a general ``no-go theorem'', aims at giving 
a precise meaning to quantization and finding maximal  
quantizable algebras.\cite{G-G-T,G-G-H,G-G,G,G-G-G}
Their definition of quantization starts with
prequantization:
modulo technicalities regarding domains of operators, 
prequantization
is a linear map from some Poisson subalgebra 
(possibly the whole algebra)
to self-adjoint operators such that the 
``Poisson bracket goes to commutator'' rule is
satisfied and the constant function $1$ is mapped 
to the identity operator.
To pass from a prequantization to a  
quantization in a general case,
the authors of Ref.~2 propose 
an approach based on the Poisson
structure and on irreducibility only: 
no relations among observables
coming from the multiplicative algebra
structure are imposed {\em a priori}. 
They look for a (minimal) ``basic'' set
carrying the kinematical information  and demand 
this set to be irreducibly 
represented. 

As a result of their work, Gotay et al.\ proposed 
a weaker than  expected
``no-go theorem'', on account of a new result which the authors
interpreted as a genuine full quantization.\cite{G-G-T,G}
The authors proved the existence of obstructions to quantization for the
sphere $S^2$ and the cylinder $T^{\ast}S^1$.\cite{G-G-H,G-G} These
results points in the same direction as Groenewold and
Van Hove's and thus could be considered as further evidence in favor 
of a strong ``no-go theorem''. However, the authors found no 
obstruction to the quantization of the torus $T^2$. In fact, it was 
shown that a prequantization of the whole Poisson algebra
of the torus produced via geometric quantization (see Ref.~1) is in
fact a  quantization in the sense defined in  
Ref.~2 and 
outlined above, meaning that a  basic set is irreducibly
represented. 
Thus, a quantization of the whole Poisson
algebra of the torus is claimed.\cite{G-G-T,G}
Although the proven irreducibility is an interesting result, we disagree that
the full prequantization of $T^2$ can be considered a physically 
acceptable quantization of the torus.
We discuss this point next, starting with a brief review of the
results presented in Ref.~10 (in what follows
$h$ is Planck's constant and $N$ a non-zero integer).

One realizes the torus $T^2$ of area
$N h$ as $\R^2/\Z^2$ with symplectic form
\be
\label{vform}
{\omega}_{{}_N} = N h\, dx\wedge dy
\ee
and identifies the Poisson algebra $C^{\infty}(T^2)$ with the 
periodic $C^{\infty}$-functions $f$ on $\R^2$ 
$$f(x+m,y+n)=f(x,y), \ \ \ \ \ \forall \, m,n\in \Z.$$
The prequantum Hilbert space is given by the rules of 
geometric quantization; it is
made of sections of a complex line bundle over $T^2$.
In  Ref.~1 it is shown
that the prequantum Hilbert space can be realized as the completion of
the space ${\Gamma}_N$ of complex $C^{\infty}$-functions  
$\f$ on $\R^2$ satisfying
\be
\label{1}
\f(x+m,y+n)=e^{2\pi i N m y}\f(x,y), \ \ \ \ \ \forall \, m,n\in \Z,
\ee
with inner product 
$$<\f,\f '>=\int_{[0,1)\times [0,1)} dxdy\,\, {\bar \f}\, \f'\ .$$
The prequantization map ${\cal P}_{{}_N}$ is given by
\be
\label{2}
{\cal P}_{{}_N}(f)\f=f\f-{i \over 2\pi N} \left({\partial f \over \partial x}
\left({\partial \f
 \over \partial y}
-2\pi N i x \f\right)-{\partial f \over \partial y}{\partial \f \over 
\partial x}\right)\, ,
\ \ \ \ \forall f \in C^{\infty}(T^2).
\ee
In the $N=1$ case (the one considered in Ref.~2 and Ref.~10),
this representation can also be realized on $L^2(\R)$, via the isomorphism
$W$ (defined between the dense subsets 
${\Gamma}_1$ and ${\cal S}(\R)$, the Schwartz space) given by
\be
\label{3}
(W\psi)(x,y)=\sum_{k=-\infty}^{+\infty} \psi(x+k)\, e^{-2\pi i k y}
\ee
\be
\label{4}
(W^{-1}\f)(x)=\int_0^1dy\, \f(x,y),\ \ \ \ \ \f\in{\Gamma}_1,\, 
\psi\in {\cal S}(\R)
\ee
Under this transformation the operators $({\partial \over \partial y}
-2\pi i x)$ and ${\partial \over \partial x}$ go over to the operators
$-2\pi i x$ and ${d \over dx}$, respectively.

For the $N=1$ case, Gotay considers the most natural 
basic set of functions which globally separates points, namely the set
\be
\label{bset}
\{\sin (2\pi x),\cos (2 \pi x),\sin (2\pi y),\cos (2 \pi y)\}
\ee
and then show that it is irreducibly represented. 
One is, thus, faced with a full quantization
of the torus, by Gotay et al.\ criteria. 
This situation is presented as a genuine
``go theorem'', and used in 
Ref.~2 to help formulating
the hypothesis of a possible general ``no-go theorem''.

We believe, however, that this representation 
of the Poisson algebra of the
torus cannot be considered a physically meaningful quantization of the
torus. Consider, for instance, 
the quantum operators representing functions depending on $x$ only, in
the $L^2(\R)$ representation:
\be
\label{5}
\big({\cal P}_{{}_1}(f)\psi\big)
(x)=\left(f(x)-x{d\over dx}f(x)\right)\psi(x)\, .
\ee
The form of the operators clearly shows that they are unbounded, having
as spectrum the whole real line.
This drastic enlargement of the spectrum of classically bounded observables
has, of course, dramatic consequences: for instance, the quantum
measurement of the mean value of $\sin (2 \pi x)$ 
(or $\sin^2 (2 \pi x)$) can produce any
real value. The operator $\big({\cal P}_{{}_1}(\sin (2 \pi x))\big)^2$
is positive of course, although not bounded, but the quantum observable
${\cal P}_{{}_1}\big(\sin^2 (2 \pi x)\big)$ is neither.
This is an example of a general fact, which is the disappearance 
at the quantum level of relations among pairs of
commuting classical observables $f(x)$, $g(x)$. In particular, for the above 
mentioned operators one gets:
\be
\label{6}
\big({\cal P}_{{}_1}(\sin (2 \pi x))\big)^2-
{\cal P}_{{}_1}\big(\sin^2 (2 \pi x)\big)
=4\pi x^2\cos^2(2\pi x)\, ,
\ee
where it is clear that the multiplicative operator appearing in the
right hand side does not correspond to the quantization of any
observable on the torus. If this representation were to be 
physically realizable,
one could not infer the observable values of 
${\cal P}_{{}_1}\big(\sin^2 (2 \pi x)\big)$ just by measuring 
${\cal P}_{{}_1}(\sin (2 \pi x))$. 

Among the broken classical relations one finds also the trigonometric one, 
which appears strongly violated at the quantum level:
\be
\label{7}
\big({\cal P}_{{}_1}(\sin (2\pi x))\big)^2+
\big({\cal P}_{{}_1}(\cos (2\pi x))\big)^2=
1+4{\pi}^2x^2\, ,
\ee
where the unobservable operator $x$ again appears. 
Therefore, and although no new local degrees of freedom entered the
quantum theory, we find no traces
of the global aspects of the phase space one has begun with.

\section{Discussion}
\noindent
It is clear from the unboundedness of the spectrum of the 
quantum operators mentioned in Sec.~2 and the related violation 
of the trigonometric relation
that information about the global topology of the phase space $T^2$ was lost.
This representation effectively ``decompactifies'' the phase space. 
This ``decompactification''
accounts for the infinite dimensionality of the Hilbert space,
which goes against the physical expectation of having only a finite number
of independent quantum states associated with a compact phase space.

In our opinion,
this example shows that irreducibility by itself does not 
guarantee, in a  general symplectic manifold,
that one gets a physically acceptable quantum theory.
In globally non-trivial cases
functional relations among basic observables appear as  essential,
since they carry information about the global topology of the 
phase space. 
In the present case of the torus we showed that 
crucial global topological 
information was indeed lost in the quantization process, although
the irreducibility requirement was satisfied.
 
It is worth mentioning that the Dirac-like approach based on the 
strict ``Poisson bracket goes to commutator'' rule does not look appropriate
in the torus case. Contrary to other known cases, e.g. cotangent 
bundles and $S^2$, the Poisson algebra of the torus does not seem to
admit subalgebras that separate points (and contains the constant function $1$)
, other than $C^{\infty}(T^2)$ itself
(and dense subalgebras, e.g the one generated by the basic set (\ref{bset})
and the constant function $1$, which is made of functions having a finite 
Fourier decomposition). 
In particular, it is known that no such
finite-dimensional subalgebra can be found.\cite{G-G-G}
Thus, regarding question (a) in the introduction, it seems that for $T^2$ 
one is bound to impose the Dirac quantum condition on the whole 
Poisson algebra, like Gotay et al.\ did.
But it was recently proved that no 
non-trivial finite-dimensional
Lie representation of $C^{\infty}({\cal M})$ can be found, for any connected
compact symplectic manifold $\cal M$.\cite{8} 
Also, it is an established fact that for compact $\cal M$
every non-trivial prequantization of $C^{\infty}({\cal M})$ in a
infinite dimensional space includes unbounded operators.\cite{Av}
Thus, it seems that, just like in Gotay's proposal, every conceivable
Dirac-like  quantization of the torus will produce unbounded operators,
thus leading to problems like the ones we discussed.

On the other hand it is known that a sequence of finite-dimensional
Lie algebras {\it g}($N$) exists, such that, in the $N\to\infty$
limit a representation of the Poisson algebra $C^{\infty}(T^2)$ 
is approached.
Thus, although no {\it g}($N$) is a representation
of $C^{\infty}(T^2)$ (they are in fact representations of a deformed algebra),
the correct classical limit is achieved.\cite{9,10,11,12} 

\nonumsection{Acknowledgments}
\noindent
I thank Jos\'e Mour\~ao for enlightening discussions and 
constant encouragement. I thank Nenad Manojlovic and Yuri Kubyshin for
valuable discussions. This work was supported by FCT under grants 
Praxis XXI BD 3429 and CERN/S/FAE/1111/96.

\nonumsection{References}

\nonumsection{}
\noindent E-mail adress: jvelhi@ualg.pt


\begin{thebibliography}{9999}




\bibitem{Ki} A.A.~Kirillov, in {\bibit Dynamical Systems IV:
Symplectic Geometry and its Applications},
Encyclopedia Math. Sci. IV,  eds. V.I.~Arnol'd and S.P.~Novikov (Springer,
New-York, 1990) pp. 137-172.

\bibitem{G-G-T} M.J.~Gotay, H.B.~Grundling, G.M.~Tuynman,
{\bibit J. Nonlinear Sci.} {\bf 6}, 469 (1996).

\bibitem{As2} A.~Ashtekar, {\bibit Lectures on Non-Perturbative Canonical
Quantum Gravity}\/ (World Scientific, Singapore, 1991).

\bibitem{G-G-H} M.J.~Gotay, H.~Grundling, C.A.~Hurst, 
{\bibit Trans. Am. Math. Soc.} {\bf 348}, 1579 (1996).

\bibitem{G-G} M.J.~Gotay, H.B.~Grundling, {\bibit On Quantizing} $T^{\ast}S^1$,
preprint quant-ph/9609025 (1996).

\bibitem{Z} F.~Ziegler, {\bibit Quantum representations and the orbit
method}, in {\bibit M\'ethode des orbites et representations quantiques},
PhD. Thesis (Universit\'e de Provence, 1996)


\bibitem{As-T} A.~Ashtekar, R.~Tate, {\bibit J. Math. Phys.} {\bf 35},
6434 (1994)

\bibitem{A-I} A.~Ashtekar, C.J.~Isham, {\bibit Class. Quant. Grav.} 
{\bf 9}, 1433 (1992)

\bibitem{A-L} A.~Ashtekar, J.~Lewandowski, in {\bibit Knots and quantum 
gravity}, eds. J.~Baez (Oxford university Press, Oxford 1994) 


\bibitem{G} M.J.~Gotay, in
{\bibit Quantization, Coherent States and Complex Structures}, 
eds. J.-P.~Antoine
{\it et al.} (Plenum, New York, 1995) pp. 55-62.


\bibitem{G-G-G} M.J.~Gotay, J.~Grabowski, H.B.~Grundling,
{\bibit An Obstruction to Quantizing Compact Symplectic Manifolds},
preprint dg-ga/9706001 (1997)

\bibitem{8} V.L. Ginzburg, R. Montgomery, {\bibit Geometric Quantization 
and No Go Theorems}, preprint dg-ga/9703010 (1997).

\bibitem{Av} A. Avez, {\bibit C.R. Acad. Sci. Paris}\/ {\bf A279}, 785 (1974)

\bibitem{9} M.A. Rieffel, {\bibit Commun. Math. Phys.} {\bf 122}, 531 (1989)

\bibitem{10} D.B. Fairlie, P. Fletcher, C.K. Zachos, 
{\bibit Phys. Lett.} {\bf B218}, 203 (1989)

\bibitem{11} M. Bordemann, E. Meinrenken, M. Schlichenmaier,
{\bibit Commun. Math. Phys.} {\bf 165}, 281 (1994)

\bibitem{12} V. Aldaya, M. Calixto, J. Guerrero, {\bibit Commun. Math. Phys.} 
{\bf 178}, 399 (1996)

\end{thebibliography}
\end{document}